\documentclass[prb,twocolumn,amsmath]{revtex4}

\usepackage{graphicx}

\begin{document}

\title{Comments On Supersolidity}
\author{W. M. Saslow}
\email{wsaslow@tamu.edu}
\affiliation{ Department of Physics, Texas A\&M University, College Station, Texas 77843-4242}

\date{\today}

\begin{abstract}
Assuming that the well-confirmed non-classical rotational inertia (NCRI) effect in solid $^4$He, suggested by Leggett, indicates supersolid behavior, we make a number of remarks about both theory and experiment.  (1) The long-wavelength, low-frequency (``hydrodynamic'') part of the theory of Andreev and Lifshitz has nine variables, and thus must have nine modes.  We find a new mode associated with lattice point diffusion (and thus vacancy diffusion); it may explain the absence of supersolid behavior in low-frequency pressure-driven flow. (2) The observed upper limit for the NCRI fraction (NCRIf) of about 20\%, in disordered samples, is more-or-less the same as the already predicted upper limit for the superfluid fraction of a well-ordered crystal; we argue that this may not be a coincidence.  (3) The negative experimental evidence for a second propagating hydrodynamic mode (expected to be fourth sound-like) may be due to the long relaxation times $\tau$ at low temperature $T$; only for frequencies satisfying $\omega\tau\ll1$ does the hydrodynamic theory apply.  (4) The fundamental principles of quantum mechanics imply that Bose-Einstein condensation is not necessary to define a quantum-mechanical phase; therefore the absence of a finite condensate fraction $f_{0}$ does not necessarily imply the absence of superfluidity. (5) Just as vortices should avoid occupied lattice sites to provide a vortex-lattice interaction, the lattice should interact with the vortices to provide a lattice-vortex interaction; thus dislocations should interact with vortices, whose motion is affected by rotation.  A relatively strong vortex-lattice interaction should also occur for superconductors with short coherence lengths, making both solid $^4$He and some high temperature superconductors candidates for the hypothesised vortex liquid phase.  The vortex-lattice interaction also may be relevant to the observation of small changes in the shear response at $T_{c}$, and to complex observed hysteresis and relaxation effects. 

\end{abstract}
\pacs{67.80.kb,61.72.Hh, 62.40.+i}

\maketitle

\section{Introduction}

Following the ac torsion oscillator (TO) experiments of Kim and Chan,\cite{KC1,KC2} which show that solid $^4$He displays a non-classical rotational inertia (NCRI), a number of groups have independently observed ac NCRI.\cite{RR1, Shira1,Koj1,Penzev}  Further, Rittner and Reppy have shown that the ac NCRI fraction (NCRIf) can be suppressed considerably (but not completely) by making good crystals;\cite{RR2} and that the NCRIf can be enhanced to as much as 20\% if the lattice disorder, caused by nonuniform crystal growth, is great enough.\cite{RR3}  Small amounts of $^3$He impurities also produce disorder, but do not contribute as significantly to the NCRI as lattice disorder.  Many microscopic theories, which calculate the superfluid fraction $f_{s}$ and equate it to the NCRIf, are challenged by this value of 20\%.  


Were it not for the fact that the NCRI effect in solid $^4$He is observed repeatedly by laboratories throughout the world, it would be easy to dismiss the possibility of supersolid behavior.  After all: 

(1) the apparent thermodynamic signature of a phase transition is clear but not sharp,\cite{Lin07} and there apparently is missing entropy that should be associated with such a phase transition;\cite{Balatsky07} 

(2) there is no theory for the effect of $^3$He impurities on the $T_{c}$ and value of the NCRI;\cite{Kim08}

(3) there is no unambiguous observation of superflow in non-NCRI experiments;\cite{Day05,Day06,Ray08,RittnerChoi} 

(4) there is no indication why the NCRI seems to saturate at around 20\%, and so far will go no higher.\cite{RR3} 

(5) there is to date no experimental indication of a second long wavelength propagating mode;\cite{Aoki08} 

(6) there is to date no experimental indication of a finite condensate fraction.\cite{Diallo07} 

(7) there are unusual hysteresis and relaxation effects.\cite{AokiPRL08,ClarkMaynardChan08, DavisScience09,PenzevYasutaKubota08}

(8) there is the puzzling observation that the static shear modulus tracks with the NCRI, stiffening at lower temperatures.\cite{DayBeamishNature07}

The present work addresses the last six issues.  

As to (3), Sect.~II notes that two types of superflow have been discussed theoretically, one based on vacancies (both by Andreev and Lifshitz\cite{A&L} and by Chester\cite{Chester}) and one based on more conventional superflow, 
by Leggett.\cite{Leggett1}  We show that buried within the Andreev and Lifshitz theory is the prediction of an additional mode (not remarked upon previously) associated with the density of lattice points but not with normal fluid or superfluid motion, and that it is diffusive.  (A similar mode was noted for ordinary solids by Fleming and Cohen.\cite{FlemCo})  We interpret this mode to mean that vacancy flow is diffusive.  Since the theory of Leggett is the only one for which NCRI was discussed,\cite{Leggett1}  we take the experiments to indicate that the type of superflow considered by Leggett, but not that considered by Chester or by Andreev and Lifshitz, is taking place.  Nevertheless, we believe that the Andreev and Lifshitz theory does correctly describe the time-dependent response of a superfluid of the type envisioned by Leggett.  We do not consider the possibility that the apparent superflow is due to thin liquid films along grain boundaries,\cite{BalScience06} since Ref.~\onlinecite{ClarkWestChan07} observes NCRI in large crystals of $^4$He.

As to (4), Sect.~III  argues that calculations already performed, which yield an upper limit of 20\% for the superfluid fraction,\cite{Saslow05} may be relevant to the experiments on highly disordered (perhaps glassy) crystals, despite the fact that the theory was developed for a crystalline solid.\cite{Saslow76}  Moreover, recent work shows that the phase function, which defines the superfluid velocity, may depend on multiple particle coordinates, so that the concept of the phase function should be expanded.\cite{SasGalRe}

As to (5), Sect.~IV argues that the hydrodynamics must be reconsidered, with an additional velocity $\vec{v}_{L}$ for the lattice, which at $T=0$ behaves similarly to a porous medium that encloses a dilute superfluid.  In this case the elastic modes and the expected new superfluid mode may lie very close in velocity, since the supersolid and solid elastic responses may be very similar.  Moreover, the temperature $T$ may be so low that equilibrium-producing collisions are too infrequent for the hydrodynamic theory to generate the second mode (expected to be fourth sound-like) associated with superfluidity.\cite{A&L}  

As to (6), Sect.~V argues that a quantum-mechanical state does not require a (relatively easily calculated) finite condensate fraction $f_0$ in order for a system to have an overall phase that defines superflow.  
(In other words, we argue that Bose-Einstein condensation is not necessary for superfluidity.)  Indeed, the fundamental principles of quantum mechanics imply that for a pure quantum state the phase function, whose gradient yields the superfluid velocity, is uniquely defined (up to an arbitrary constant), and therefore that it is unnecessary to have a finite condensate fraction to define that phase function.  Hence, although a finite condensate fraction makes it easier to identify the phase function, and thus to assure superfluidity (subject to the Landau criterion), its absence does not assure the absence of superfluidity.  We also comment on how substrate properties might affect superfluidity. 

As to (7) and (8), Sect.~VI argues that the lattice-vortex interaction is likely capable of causing unusual hysteresis and relaxation effects, and slight tracking of the static shear modulus with the NCRI.  Moreover, the presence of a vortex-lattice interaction would explain why the vortex lattice hypothesis would be appropriate for solid $^4$He and high $T_{c}$ superconductors, but not for liquid $^4$He or ordinary superconductors.  

Space limitations prevent adequate referencing to the many works in this area.  Ref.\onlinecite{BalRev} and Ref.\onlinecite{Proko} contain reviews, and Ref.\onlinecite{Boronat09} exemplifies fairly recent Monte Carlo calculations.  
We also note the informative work by Chan.\cite{ChanScience08}

Sect.~VII presents some concluding remarks.  An Appendix considers the Landau criterion applied to the excitations of a wall, concluding that, in principle, metallic container walls could destroy superfluidity, but that likely the tails of the electron wavefunctions are excluded from the solid $^4$He. 

\section{Two Types of Superflow?  The Vacancy DIffusion Mode.}

Two forms of superfluid solid have been discussed in the literature.  Andreev and Lifshitz (A\&L)\cite{A&L} (as well as Chester\cite{Chester}) considered that superfluidity would be due to very mobile vacancies.  A\&L give both a tight-binding-like quantum-mechanical theory of defect motion and a long-wavelength, low-frequency (hydrodynamic) theory of a superfluid solid.  For the latter A\&L employ a lattice displacement $\vec{u}$ to define the strain, and note that their fourth sound-like mode at temperature $T=0$ ``represent[s] oscillations of the crystal density with fixed lattice sites ($u_{i}=0$).  One can also show that these correspond to oscillations of the density of defectons.''  On the other hand, Leggett, who presented the only $T=0$ microscopic theory for the $f_s$ of a supersolid (properly, an upper limit on $f_s$),\cite{Leggett1} considered that the density is fixed, and that the flow occurs because of a phase gradient (what we have called ``phase flow''\cite{Saslow76}).  Physicists are familiar with phase flow in the context of charged supersolids -- superconductors -- although they tend to think of superconductors as a charged fluid.  Nevertheless, the electrons in a superconductor live within a lattice, and have wavefunctions that reflect the periodicity of the lattice.  Because the electronic states are relatively delocalized, their superflow patterns yield negligible suppression of the superfluid fraction, as discussed in Ref.\onlinecite{Saslow76}.  

We now show that there is an additional mode, not discussed by A\&L, that is present in their hydrodynamic (e.g., long-wavelength, low-frequency) theory.  Existence of such a mode can be obtained by comparing the number of variables and of modes, which must be equal.  Since $\vec{v}_{s}$ is the gradient of a phase, $\vec{v}_{s}$ represents only one variable.  A count of hydrodynamic variables then yields nine: $T$, chemical potential $\mu$, superfluid velocity $\vec{v}_{s}$, normal fluid velocity $\vec{v}_{n}$, and $\vec{u}$.  On the other hand, at $T=0$ A\&L find only eight modes, distributed in four sets of doubly-degenerate modes: a mode is missing.  

To obtain this additional mode, we consider the lattice point density $n_{0}=N_{0}/V$, where $N_{0}$ is the number of lattice points in a volume $V$.  By definition of the strain $w_{ij}=\partial_{i}u_{j}$, the variations in changes in $n_{0}$ and $w_{ii}$ are exactly related by
\begin{equation}
\delta n_{0}=-n_{0}\delta w_{ii}=-n_{0}\partial_{i}u_{i}.
\label{n_0}
\end{equation}
To study $\dot{n_{0}}\equiv dn_{0}/dt$ we employ the linearized-in-amplitude version of the second of the equations (15) of A\&L
\begin{equation}
v_{ni}-\dot{u}_{i}=-\alpha_{ik}\partial_{k}T-\beta_{jk}\partial_{i}\lambda_{kj}.
\label{vn-du/dt}
\end{equation}
We now assume a mode where $\vec{v}_{n}=\vec{v}_{s}=\vec{0}$ but $\dot{u}_{i}\ne 0$.  Moreover, we assume that the dissipative Onsager cross-coupling $\alpha_{ik}$ is negligible, and that the lattice is nearly isotropic, so that $\beta_{ik}\approx \beta\delta_{ik}$.  Then (\ref{vn-du/dt}) becomes
\begin{equation}
\dot{u}_{i}\approx\beta\partial_{i}\lambda_{kk}.
\label{du/dt}
\end{equation}
Again taking the elastic system to be isotropic, so $\delta\lambda_{kk}\approx E\delta w_{kk}$, application of $\partial_{i}$ to (\ref{du/dt}) yields
\begin{equation}
\dot{n}_{0}\approx\beta E\partial_{k}\partial_{k} n_{0}.
\label{du/dtD}
\end{equation}
This is a diffusion equation, and thus there is a motion involving the vacancies only, which leads to diffusion of the vacancies.  Of course, both first sound and fourth sound also involve motion of vacancies, by $v_{ni}\approx\dot{u}_{i}$.  A similar diffusion mode, using vacancy concentration $c$ as the variable, was found for ordinary solids.\cite{FlemCo}  Note that if the $\beta_{jk}\partial_{i}\lambda_{kj}$ term is included in the A\&L equations of motion for $\vec{v}_{s}$ and the momentum density $\vec{g}$, then $\vec{v}_{s}$ and $\vec{v}_{n}$ will be brought into motion by the vacancy motion, but only to second order in the wavevector.  Hence we may ignore $\vec{v}_{s}$ and $\vec{v}_{n}$ in discussing vacancy diffusion in the long wavelength limit.  

Note that if there are $N$ particles in $V$, for particle density $n$, then the vacancy density $n_{V}$ satisfies 
\begin{equation}
{n}_{V}=\frac{N_{0}-N}{V}={n}_{0}-{n}.
\label{n_V}
\end{equation}
Since $n=\rho/m$ ($m$ is a $^4$He mass) is proportional to the total mass density, which is conserved, this leaves the vacancy motion as sharing the conserved motion of $n$ and the non-conserved motion of $n_{0}$. 

A number of pressure-driven non-NCRI experiments have been performed, in a search for a superfluid response, but in our opinion there is no clear signature of superflow.\cite{Day05,Day06,RittnerChoi,Ray08}  To our understanding, these pressure-driven non-NCRI experiments eliminate only the vacancy flow mechanism for superflow, although they are consistent with the vacancy diffusion implied by AL's hydrodynamic (but not microscopic) theory.  They do not, however, eliminate the phase flow mechanism, which is supported by the NCRI measurements.  Note that the four sets of propagating modes given in AL are consistent with what one expects for superflow due to phase flow.  The remainder of this paper almost exclusively considers only phase flow.  


\section{Is 20\% the experimental upper limit for the NCRI fraction?}

As already noted, although clean samples of solid $^4$He exhibit very low NCRIf,\cite{RR1} of order 0.01\%, samples that are deliberately made as noncrystalline as possible show up to 20\% NCRIf.\cite{RR2,RR3}  This section argues that 20\% may be a realistic upper limit for $f_s$, based on calculations that had already been performed before the results of Refs.\onlinecite{RR2} and \onlinecite{RR3} became available. 

Leggett's work on the superfluid fraction $f_s$ of a solid actual considered only an upper limit $f^{u}_s$, by employing a variational form for the phase function;  this was implicitly taken to be one-body in nature, as for ordinary superfluids.\cite{Leggett1} 
In addition, for simplicity, the phase function was taken to depend only on the coordinate in the flow direction, making the phase function not only one-body, but also one-dimensional.\cite{Leggett1}  Ref.\onlinecite{Saslow76} extended Leggett's theory of the upper limit for $f_s$ by retaining a one-body phase function but making it three-dimensional.  Taking the one-body density to be a sum of gaussians, this work calculated an upper limit $f^{u}_s$ for $f_s$ as a function of the localization of the gaussian.\cite{Saslow76}  For relatively localized gaussians (as appropriate to a classical solid) $f^{u}_s=0$, and for relatively delocalized gaussians (as appropriate to the electrons in a superconductor) $f^{u}_s=1$.  The upper limit was estimated to lie between 5\% to 20\%, the large uncertainty due to uncertainties at that time both in the extent of localization of solid $^4$He, and computational limitations on convergence of $f^{u}_s$.  Also note Ref.~\onlinecite{FernandezPuma}, which estimated $f^{u}_{s}$ to lie between 0.2 and 0.4.

Recently both one-body, one-dimensional and one-body, three-dimensional upper limits for $f_s$ were computed, using one-body densities taken from quantum-mechanical Monte Carlo calculations.\cite{GalReSas}  The one degree-of-freedom $f^{u}_s$ for hcp solid $^4$He at $\rho=0.029$ \AA$^{-3}$ employs the average density normal to the flow direction.  This upper limit depends very much on the flow direction: flow along $z$ gave $f^{u}_s=0.384$, flow along $x$ gave $f^{u}_{s}=0.939$ and flow along $y$ gave $f^{u}_{s}=0.799$.  The three-degree-of-freedom upper limit, on the other hand, was (as expected) very nearly isotropic, with $f^{u}_s=$0.21-0.22.  Given that we expect Nature to choose the better variational wavefunction, the three-dimensional one-body phase function, yielding a 20\% upper limit, will be preferred.\cite{Saslow06}  

How to go beyond a one-body phase, to include higher-order correlations, has also been considered.  The formalism for the case of a two-body phase has been developed, but not applied, because currently the three-body correlations needed to compute the two-body phase function upper limit are unavailable.\cite{SasGalRe}  Specifically, the wavefunction with superflow for $i=1,N$ particles can take the form
\begin{eqnarray}
\Psi(\{x_{i}\})&=&\Psi_{0}(\{x_{i}\})\exp(i\Phi), \\
\Phi(\{x_{i}\})&=&\sum_{i}\phi_{1}(x_{i})+\frac{1}{2(N-1)}\sum_{ij}\phi_2(x_{i},x_{j})\cr
&&+\dots
\end{eqnarray}
Here $\Psi_{0}$ is the ground state wavefunction without flow, $\phi_{1}$ is the usual one-body phase function familiar from liquid $^4$He, and $\phi_{2}$ is a new phase function that includes two-body correlations.  The above refers, of course, to $T=0$. 

Even if the upper limit for $f_s$ is finite, to be a superfluid the system must satisfy two other criteria.  First, if the Landau criterion is not satisfied, the system will not be superfluid, no matter how complex a variational phase function is taken, because it will be energetically favorable for the flowing state to slow down by generating excitations with momentum along the flow direction.  Second, if the wavefunction is not topologically connected (except perhaps at a finite number of points), the system will not be superfluid, because (at least for a one-body phase function) it will be possible to find a continuous path along which the number density is zero but along which the phase can accumulate with no cost in additional flow energy -- and thus no superfluid fraction.\cite{Leggett1}  It is not clear how complex a variational phase function is needed to be sensitive to a wavefunction that is topologically connected in a more subtle way than can be described by one-body effects.\cite{Leggett1,Leggett3}  Presumably, if a sequence of variational phase functions with successively higher correlations (one-body, two-body, etc.) is considered,\cite{SasGalRe} then as the higher correlations are included the upper limit for $f_s$ of a topologically disconnected state will go to zero.  For a translationally-invariant fluid, for whose ground state a one-body phase function gives $f^{u}_s=1$, higher correlations have been shown not to decrease $f^{u}_s$.\cite{Leggett3,SasGalRe}  The next paragraph will invoke this result. 

Let us consider the difference between solid $^4$He in crystalline and disordered form.  Other than the disorder, the one-body densities should display similar amounts of localization.  Therefore the one-body phase function upper limit for $f_s$ of 20\% should apply to both crystalline and disordered systems.  Differences will show up primarily in correlations.  From the absence of NCRI in good crystals, we conclude that the many-body correlations make it necessary to include higher-order phase functions, which decrease the upper limit on $f_s$ to a small fraction of a percent or even zero.   How can the disordered state have an experimental NCFRf of about 20\%?  We have just noted that the liquid state cannot utilize higher-order phase functions to decrease its $T=0$ upper limit for $f^{u}_s$ below 1.\cite{Leggett3,SasGalRe}  If the disordered solid has static but liquid-like correlations, then for the disordered solid the higher order phase functions may not be able to take advantage of these correlations to cause significant further lowering of the flow energy.  This hypothesis would have to be confirmed by microscopic  calculations.  Exact calculations of $f_s$ can be made by evaluating the winding number, but require the exact thermal weighting of the exact many-body wavefunctions.\cite{PolCep84, CepRMP}  

Thus, to answer the question posed by this section, we believe that it is reasonable for the upper limit for $f_s$ to remain very near 20\%.  Under higher pressure, where greater localization is expected, one can anticipate this upper limit to decrease, but this is only a qualitative argument.  

\section{Is the Hydrodynamic Theory of a Supersolid Complete?}

As discussed in Sect.~II, Andreev and Lifshitz developed a hydrodynamic theory for a type of supersolid.  As noted, it had three velocities: $\vec{v}_{n}$, $\vec{v}_{s}$, and $\dot{\vec{u}}=\partial\vec{u}/\partial t$.  Here $\vec{v}_{n}$ and $\vec{v}_{s}$ were given Galilean transformation properties.
The associated mass densities $\rho_{n}$ and $\rho_{s}$ were taken to sum to the total mass density $\rho$.  By adding to the rate of entropy production a specific amount of a set of terms that summed to zero by a Gibbs-Duhem relation, $d\vec{u}/dt$ was made Galilean.  This theory was extended to include nonlinear terms in Ref.\onlinecite{SaslowSS}.  Liu pointed out that there could also be a ``thermal'' supersolid,\cite{LiuSS} with no obvious NCRI properties; we will assume that solid $^4$He is not described by a theory of that sort.  Although developed with vacancy flow superfluidity in mind, we believe that the essential aspects of the Andreev-Lifshitz theory apply to phase flow superfluidity. 

If superfluidity of solid $^4$He is real, with $f_s<1$ at $T=0$, and if the normal fluid fraction $f_n=0$ at $T=0$ because there are no excitations, then a new density must compensate for that missing density the difference between the total density and the $T=0$ superfluid density, as noted earlier in Ref.~\onlinecite{Saslow05}.  To this new density  an additional velocity must be associated, whose dynamics must be determined.  

Let $f_L$ denote the additional mass fraction associated with the system, and denote the additional velocity variable $\vec{v}_{L}$.  This gives the system four velocities: $\vec{v}_{n}$, $\vec{v}_{s}$, $d\vec{u}/dt$ for the lattice points, and $\vec{v}_{L}$ for the new variable, somehow associated with the lattice, but in principle distinct from $d\vec{u}/dt$.  To ensure that all four of these velocities are Galilean within the hydrodynamics approach of Andreev and Lifshitz one finds -- as they did -- that $\vec{v}_{n}\approx d\vec{u}/dt$ (neglecting gradient terms that lead to defect diffusion when $\vec{v}_{n}=\vec{0}$).  We also find that $\vec{v}_{L}$ decays toward $d\vec{u}/dt$ (presumably due to Umklapp processes).\cite{Saslow4vels}  The implication is that, at frequencies low relative to the Umklapp frequency $\tau_{U}^{-1}$, the normal modes found by Andreev and Lifshitz (elastic waves and fourth sound) will apply, but at higher frequencies the response will be more complex.  At low temperatures the condition $\omega\tau_{U}\ll1$ may require such small frequencies that the wavelengths may be larger than the system under study, and thus the modes might not be observable in an apparatus of realistic laboratory dimensions.  For such low frequencies the normal fluid density $\rho_{n}$ employed by Andreev and Lifshitz should be interpreted as including both the $\rho_{n}=f_{n}\rho$ of excitations and the density $\rho_{L}=f_{L}\rho$ associated with $\vec{v}_{L}$, as in fact seems to have been done implicitly by A\&L.  


Thus, to answer the question posed by this section, we believe that there is a need for a more complete hydrodynamic theory of supersolids.  

\section{Is a Finite Condensate Fraction Necessary for Superfluidity?}

The NCRI experiments cited above establish that non-classical rotational inertia is a robust phenomenon in solid $^4$He at low temperatures.  We believe that the most natural explanation for this phenomenon is phase flow superfluidity of the solid.  

It is commonly thought that a finite condensate fraction $f_0$ is necessary for phase flow superfluidity, although the superfluid fraction $f_s$ is what one compares with the experimental NCRI fraction.  For realistic microscopic models, vacancies seem to be present at the level of a part in ten thousand, and they may be necessary for a finite condensate fraction.\cite{Boronat09}  The present section argues that a finite condensate fraction is not necessary for superfluidity.  This would explain its failure to appear in neutron scattering experiments.\cite{Diallo07}  An equivalent statement is that Bose-Einstein condensation is unnecessary for superfluidity. 

As background we note that Landau's theory for liquid $^4$He established that a translationally invariant Bose system, to be superfluid, must be stable relative to decay of excitations with momentum in the flow direction.\cite{Landau1941}  Moreover, Landau presented a theory for quantized sound waves as those excitations, with superflow relative to a wall being stable for flow velocities below the sound velocity.  Later developments indicated that the excitation spectrum in liquid $^4$He has a more complex structure, with the ``roton'' at finite momentum being the most likely bulk excitation to destroy superflow,\cite{Landau1947,Feynman54,FeynmanCohen56} and quantized vortices generated at surfaces providing an even more effective candidate to destroy superflow.\cite{Feynman55}  Bogoliubov showed that for a weakly interacting Bose gas, where $f_0$ is significant, the low momentum excitations have a linear spectrum with sound velocity proportional to $\sqrt{f_0}$.\cite{Bogoliubov}  From this one might extrapolate that for too strong an interaction the condensate fraction might go to zero, the sound velocity might go to zero, and the system would be unstable to sound waves.  Therefore the superfluid fraction would go to zero if $f_0$ went to zero. 

Despite the fact that the Bogoliubov theory is usually not invoked for stronger interactions, it lurks in the background.  Given that modern theories of liquid $^4$He yield a finite sound velocity (otherwise the system would be thermodynamically unstable, in the sense of a second order transition, to density fluctuations), one can seriously question whether one can extrapolate the Bogoliubov theory to the limit of total depletion and therefore sound velocity equal to zero.  Moreover, numerous theories of boson quantum hydrodynamics\cite{Sunakawa,Morrison,Grest} obtain the sound velocity with the full density $\rho$ in place of the condensate density $f_{0}\rho$.  In what follows, the Bogoliubov theory plays no role.  However, since our conclusion is at variance with the above extrapolation of Bogoliubov theory, we wish to head off that line of argumentation.  

Consider a system of any number of identical Bose particles. In the absence of interactions, the eigenstates of the system as a whole are products of eigenstates of the individual particles.  The Landau argument applied to the extremely low energy-per-momentum free-particle excitations indicates that the system would not be superfluid.\cite{Landau1941}  Now turn on the interactions.  By the fundamental principles of quantum mechanics, any eigenstate of this system can be written as a sum of configurations, each configuration being a product state for the non-interacting system.  By the diagonalization procedure associated with finding eigenstates of the Schr\"odinger equation, each configuration has a well-defined amplitude and phase relative to every other configuration.  Only the overall phase of the system is unknown.  This is true even if the condensate fraction $f_0$ goes to zero.  (It is also true for fermions, and for arbitrary mixtures of particles of different types with different statistics.)  Thus an overall phase for the system can be defined even without a finite $f_0$.  Eq.~(1) indicates how the phase $\Psi$ can appear, and it does not require that the ground state $\Psi_{0}$ have a condensate. 

Here is an analogy that we think captures the essence of the no-condensate physics.  Corresponding to the noninteracting system, with $f_0=1$, let us take Rome without any other cities.  Corresponding to the state that develops for weak interactions  -- a superposition of noninteracting eigenstates with correlations most easily described relative to the condensate, and with $f_0<1$ -- let us take the Roman empire in its beginning stages, with cities whose only inter-city roads lead to Rome.  Next, corresponding to the state that develops when the interactions are so strong that the condensate is significantly depleted ($f_0\ll 1$), let us take the case where the Roman ``suburbs'' like Paris and Milano develop roads connecting one another.  Finally, corresponding to the state that develops when the interactions are so strong that the system has no finite condensate ($f_0=1$), let us take to correspond to the case where the inter-city roads to Rome become no more important than any other inter-city roads (no offense to our Roman colleagues).  Just as the different configurations correlate just as strongly to one another as to the condensate, in this analogy the suburbs communicate as strongly with one another as with Rome.  

Let us return to this completely condensate-depleted interacting system of bosons, which could be solid or liquid, but for the moment assume only a condensate-depleted liquid.  Let such a system be confined in an annulus that is ``at rest'', and consider the ground state of the system.  If the system is brought into uniform motion relative to the annulus, then a ``boosted'' version of the ground state will be the lowest state of energy unless the Landau criterion is violated or if the wavefunction is topologically disconnected.  To our understanding, the considerations of Ref.\onlinecite{Leggett1} would then apply, and therefore that work does not require a finite condensate. 


Thus, to answer the question posed by the title of this section, a finite $f_0$ likely is not necessary for superfluidity.  An Appendix has further considerations on the Landau criterion applied to superfluids in containers with conducting walls. 

\section{Vortex-Lattice Interaction}
We have previously noted\cite{Saslow05} that vortices can be expected to avoid the lattice sites, thus leading to a potential landscape with preferred regions for vortices (both lines and rings) in the solid.  (One can imagine a repulsive energy landscape for the vortex-lattice interaction that mimics the density profile of the lattice, as in a contact interaction.)  We now note that, just as vortices see the lattice, the lattice sees the vortices.  Therefore processes involving dislocations (and vacancies) may be sensitive to the presence of vortices.  We believe that this may be relevant to a number of experiments, which we discuss in the present section. 

We have assumed that the observed NCRI is due to phase flow superflow, following Leggett.\cite{Leggett1}  One important work contrary to this is the observation that the static shear modulus tracks with the NCRIf, stiffening at lower temperatures.\cite{DayBeamishNature07}  It was proposed that such stiffening is due to dislocations losing their ability to respond to transverse stress.\cite{DayBeamishNature07}  
However, it has been noted that this effect is too small to account for the frequency upshift observed in TO's.\cite{ChanScience08}  If vortices are present (even when the system is not rotating) they would be able to interact with the dislocations, and some of the pinning they see to the lattice may translate to an additional pinning by the dislocations.  This would explain the puzzling shear modulus effect as a secondary, but related, phenomenon. 

The possibility of a vortex liquid (VL) in solid $^4$He and in high $T_{c}$ superconductors for some intermediate temperature range\cite{PWANP07,PWAPRL08} is one for which the vortex-lattice interaction is relevant.  (In a VL state there is local, but not global order.)  An analogy is made between rotation frequency in superfluids and magnetic field in superconductors.  (Because one is ac and the other dc, one should be careful in comparing ac NCRI in solid $^4$He and dc susceptibility in superconductors.)  An apparent difficulty with this analogy is that in liquid $^4$He there is a low temperature superfluid state but (apparently) no vortex liquid state.  However, the presence of the interaction between vortices and the lattice in solid $^4$He makes it differ from liquid $^4$He, so the absence of a VL in the liquid is not an argument against it in the solid.  Ref.~\onlinecite{Penzev09} interprets their NCRIf data to support a vortex liquid to supersolid transition as $T$ is lowered.  

For solid $^4$He one can expect the vortex-lattice interaction to be relatively strong, because the vortex core is relatively small, and does little spatial averaging over lattice inhomogeneities.  For superconductors the vortex core size is comparable to the coherence length.  For ordinary superconductors the coherence length is large compared to the lattice constant; therefore lattice pinning (rather than pinning due to extended defect regions) is expected to be small.  However, superconductors with small coherence lengths (such as the cuprates, for which the VL phase was proposed\cite{PWANP07,PWAPRL08}) should have a larger vortex-lattice interaction, for which the vortex liquid model is more appropriate.  The presence of this interaction, and the consequent barriers to vortex flow, may suppress vortex recombination, and thus extend the vortex regime to higher temperatures than otherwise. 





Refs.~\onlinecite{AokiPRL08,ClarkMaynardChan08,DavisScience09} study relaxation associated with ac NCRI.  Ref.~\onlinecite{PenzevYasutaKubota08} performs a related study of hysteresis in ac NCRI.  Note that Ref.~\onlinecite{AokiPRL08} studies large velocities $v$ relative to the critical velocity $v_{c}$, Ref.~\onlinecite{DavisScience09} studies relatively small $v$, and Ref.~\onlinecite{ClarkMaynardChan08} considers both regimes.  
Note that vortices are driven in and out of the system in an ac NCRI experiment if $v>v_{c}$.  Given the expected interaction between vortices and lattice defects, this suggests that there might be four types of relaxation in NCRI experiments, according to whether $v> v_{c}$ or $v< v_{c}$, and whether it is the vortices or the dislocations that are relaxing (to make no mention of the vacancies).  We also note that in magnetic systems, large magnetic fields and higher temperature decrease the number of phase space regions available to a system, and lead to less hysteresis.  The same may happen with ac NCRI experiments.  This may explain the difference between high $v$ experiments at low and high $T$; at low $T$ the system is stuck within a local well that it cannot escape from thermally.\cite{AokiPRL08}  One might interpret the observed short-time exponentials\cite{AokiPRL08,DavisScience09} as due to vortex equilibration with the dislocations essentially fixed, and the longer-time decays as due to equilibration of dislocations that are subject to a ``dressing'' of vortices.  However, other possibilities might be more appropriate.

Finally we note a work that only indirectly deals with the issue of vortices and dislocations.  Ref.~\onlinecite{LT25Rittner} measures pressure in confined helium samples, finding a $T^2$ term that indicates disorder, and which can be decreased by annealing.  The authors also study pressure relaxation, which has a characteristic time that increases with decreasing temperature, and which is associated with a decreased $T^2$ term.  The annealing may be associated with vacancy diffusion.  For earlier work see Ref.~\onlinecite{Day05} and Ref.~\onlinecite{Grigorev07}.  Related to disorder, note the superglass model of Ref.~\onlinecite{MCsuperglass}, and the phenomenological response function analysis of Ref.\onlinecite{Nussinov07}.

\section{Concluding Remarks}

We would like to repeat the suggestion that ion flow in solid $^4$He might be used to generate and detect vortex rings.\cite{Saslow05, RayfieldReif}  This could be interesting for itself, but also because, for the vortex liquid state (but not the presumed lower temperature supersolid state) the relatively small vortex rings might persist better than the relatively large vortex lines.  Moreover, as noted by Leggett,\cite{Leggett04} it would be valuable to perform the Hess-Fairbank experiment,\cite{HessFairbank} whereby solid $^4$He at high temperature is rotated slowly (below any critical velocity for the formation of vorticity) and cooled into the supersolid regime.  If the system is supersolid, then the rotational inertia will decrease as the temperature is lowered.  This is a dc NCRI.  It requires that the rotational frequency $\omega$ be less than the critical frequency $\omega_{c}(T)$ for the entry of vorticity.  Note that since $\omega_{c}\rightarrow0$ as $T\rightarrow T_{c}$, very close to $T_{c}$ the measurements will not be in equilibrium, and that for fixed $\omega$, as the sample is cooled one goes from a metastable state where $\omega>\omega_{c}$ to an equilibrium state where $\omega<\omega_{c}$.  

\smallskip
\begin{acknowledgements}
I would like to thank Alexander Balatsky, Zohar Nussinov, and Haruo Kojima for commenting on the manuscript, and for valuable conversations and communications.  Valery Pokrovsky reminded me that vortex-lattice pinning should be most important for superconductors with short coherence lengths.  This work was partially supported by the Department of Energy through grant DE-FG02-06ER46278.  
\end{acknowledgements}

\appendix*
\section{Can Wall Excitations Destroy Superflow?}
We take this opportunity to comment on the Landau criterion, which assumes that the confining walls  provide the interaction that causes the superflow-destroying excitations within the superfluid.  In practice, vorticity entering at the walls, rather than ordinary non-topological excitations, destroys superfluid flow.  
However, now consider the possibility that excitations of the wall (rather than the fluid) can take up momentum, and thereby can destroy superfluidity.  The phonons of the wall material, with a finite velocity, will be stable under the Landau criterion.  Since phonons are the only low-energy excitations for insulators, the latter are stable under the Landau criterion.  However, under many circumstances the walls confining the $^4$He are made of conductors, for which quasiparticles are filled to the top of the Fermi sea.  For flow along, say $z$, an electron at the Fermi level with momentum along $x$ can absorb momentum along $z$, exciting it to above the Fermi level at the cost of very little energy.  Therefore from energetics alone it appears that interactions with conducting walls can, in principle, destroy superflow.  We are aware of no evidence for this; perhaps the effect is too small to be observed because the electron wavefunctions are exponentially suppressed from ``leaking'' into the $^4$He.  


{}

\end{document}